\documentstyle [12pt,a4,bezier] {article}
\linethickness{0.4pt}
\newcommand {\be}{\begin{equation}}
\newcommand {\ee}{\end{equation}}
\newcommand {\bc}{\begin{center}}
\newcommand {\ec}{\end{center} \nopagebreak}
\newcommand {\bl}{\begin{list}}
\newcommand {\el}{\end{list}}
\newcommand {\bea}{\begin{eqnarray}}
\newcommand {\eea}{\end{eqnarray}}
\newcommand {\C}{\mbox{\bf C}}
\newcommand {\N}{\mbox{\bf N}}
\newcommand {\Z}{\mbox{\bf Z}}
\newcommand {\x}{\mbox{\bf x}}
\newcommand {\ad}{\mbox{ad}}
\newcommand {\coadd}{\rlap{\mbox{$\Delta$}}_{\, +}}
\begin {document}
\begin {center} {\LARGE\bf The Quantum Orthogonal Mystery}\\
\vspace{.3in}
{\large A. Sudbery}\\
{\em Department of Mathematics, University of York, Heslington, York,
England YO1 5DD}\\[5mm]
Based on a talk at the XXX Karpacz Winter School, February 1994\\[1cm]
QUANTUM LIE ALGEBRAS\\
\ec

       The familiar examples of quantum groups arise from semisimple Lie
algebras, and it ought to be possible to consider them as similar sorts
of objects \cite{MLie}. Usually the emphasis is on their resemblance to
enveloping
algebras, i.e. infinite-dimensional associative algebras, but there is
always a finite-dimensional Lie algebra-like structure to be found in
a quantised enveloping algebra. Let $L$ be a simple finite-dimensional
Lie algebra, and let $U_{q}(L)$ be a quantisation of its enveloping algebra
$U(L)$. Then the representations of $U_{q}(L)$ are deformations of the
representations of $U(L)$ (and therefore of those of $L$): for every
representation $\rho$ of $U(L)$ on a finite-dimensional vector space
$V_{\rho}$ there is a representation
$\rho_{q}$ of $U_{q}(L)$ on the same vector space. Moreover, $U_{q}(L)$
itself under the adjoint action decomposes into irreducibles in essentially
the same
way as $U(L)$ \cite{JL}. In particular, it contains a copy of the deformation
of the adjoint representation of $L$; hence there is a subspace $L_{q}
\subset U_{q}(L)$ which is invariant under ad$X$ for each $X \in U_{q}(L)$.
Thus we can define a bracket on $L_{q}$ by
\be
[X,Y]=\mbox{ad}X(Y)=\sum X_{(1)}Y \kappa (X_{(2)})  \label{eq: adbrac}
\ee
where, as usual, the coproduct in $U_q(L)$ is denoted by $\Delta (X)=
\sum X_{(1)} \otimes X_{(2)}$, and $\kappa$ denotes the antipode in $U_q(L)$.
For example, in the simplest case of $L=${\bf sl}(2) the quantised enveloping
algebra has a three-dimensional subspace spanned by $V_{+}$, $V_{0}$ and
$V_{-}$ which is closed under the above bracket, as follows:
\be
\begin{array}{lll}
{[}V_{+},V_{+}]=0, & \quad \quad [V_{+},V_{0}]=-q^{-1}V_{+},
& \quad \quad [V_{+},V_{-}]=V_{0}, \\*[2pt]
{[}V_{0},V_{+}]=qV_{+},  & \quad \quad [V_{0},V_{0}]=(q-q^{-1})V_{0},
& \quad \quad  [V_{0},V_{-}]=-q^{-1}V_{-}, \\*[2pt]
{[}V_{-},V_{+}]=-V_{0},  & \quad \quad [V_{-},V_{0}]=qV_{-}, &
\quad \quad [V_{-}, V_{-}]=0.
\end{array}
\label{eq:3-brackets} \ee

 In any Hopf algebra the bracket defined by (\ref{eq: adbrac}) obeys two
versions of the Jacobi identity:
\bea
[X,[Y,Z]] &=& \sum [[X_{(1)},Y],[X_{(2)},Z]]\\
 \mbox{and}\hspace{1cm}  [[X,Y],Z] &=& \sum[X_{(1)},[Y,[\kappa (X_{(2)}),Z]]]
\eea
The first of these says that each ad$X$ is a generalised derivation of
the non-associative algebra structure defined by the bracket; the second
seems to be trying to say that the map $X \mapsto \ad X$ is, in some sense,
a representation of this algebra structure. These identities provide good
reason for thinking of the bracket (\ref{eq: adbrac}) as defining a kind
of Lie algebra. However, subspaces which are closed under the bracket
(such as the three-dimensional subspace exhibited in (\ref{eq:3-brackets}))
cannot be regarded as instances of this kind of Lie algebra, since they
are not usually subcoalgebras and so the coproduct (and therefore the
bracket) cannot be defined in an intrinsic way. Moreover, there does not
seem to be any version of the antisymmetry property of the Lie bracket
in this context.

 Woronowicz \cite{W}, working in a more geometrical framework, has produced
a different set of axioms for a quantum Lie algebra; these arise from
the properties of vector fields in non-commutative geometry. They have
only one Jacobi identity (corresponding to the representation property
of the adjoint map), but they also have an anti-commutativity axiom and
a natural notion of representation, and they have a wealth of good
finite-dimensional
examples. They are obtained from the differential geometry of quantum
groups in a similar way to the construction of classical Lie algebras
from Lie groups, and at first sight Woronowicz's theory promises to give
genuine deformations of all classical Lie algebras; in particular, it
looks as if these quantum Lie algebras should have the same dimensions
as the classical ones. But this expectation is to be disappointed; it
seems to be difficult, in the majority of cases, to construct non-commutative
differential spaces whose tangent spaces have the same dimension as in
the classical theory. Satisfactory deformations exist in the case of the
general linear groups but not, it appears, in other cases.

 In this talk I will examine the orthogonal groups to try to understand
this difference between the classical and the non-commutative geometry.
I will start by presenting Woronowicz's theory in a formulation which
emphasises its links with classical differential geometry.

\vspace{5 mm}
\begin{center}
NON-COMMUTATIVE DIFFERENTIATION
\end{center}

The differential geometry associated with a non-commutative space
(= non-commutative algebra, regarded as an algebra of functions)
in the theories of Woronowicz
\cite{W}, Wess and Zumino \cite{WZ} and their co-workers
\cite{SWZ} is based on an algebra of differential
forms on the space, defined by means of their commutation relations with
the non-commuting coordinate functions and by the algebraic properties
of an exterior derivative $d$; apart from these algebraic properties,
there is little connection with the intuitive ideas of differentiation.
Classically, differential forms are defined more geometrically in terms
of tangent vectors, which in turn are defined in terms of directional
derivatives of functions. It is possible to follow this more geometrical
line in the non-commutative case also, using ideas of Majid \cite{Mdiff}
which show something like a true process of differentiation behind the
non-commutative differential algebra. Here I will show how these ideas
lead to the relations between coordinates and differentials which are
a starting point for Woronowicz and his followers.

      Let ${\cal A}$ be a (non-commutative) algebra generated by $x^{1},
\ldots
, x^{n}$; we think of these as coordinates on a vector space $V$, and
therefore
use $V^{*}$ to denote the vector space spanned by the $x^{i}$. The additive
structure on a classical vector space $V$, which is a map from $V \times
V$ to $V$, can be expressed in terms of the algebra ${\cal A}$ of functions
on
the space by a {\em coaddition} map $\coadd : {\cal A} \rightarrow {\cal
A} \otimes {\cal A}$ \cite{Mmat, Mbraid, Vlad};
classically,
$\coadd(x^{i})=x^{i} \otimes 1 + 1 \otimes x^{i}$ and $\coadd$ is an algebra
homomorphism from ${\cal A}$ to ${\cal A} \otimes {\cal A}$ with the usual
algebra structure
in the tensor product, so that $x^{i} \otimes 1$ commutes with $1 \otimes
x^{i}$. Now suppose that ${\cal A}$ is non-commutative, with relations
between
the generators $x^{i}$ given by an $R$-matrix in the form
\be R^{ij}_{kl}x^{k}x^{l}=qx^{i}x^{j}
\ee
where $q$ is an eigenvalue of $R$. Then the $\coadd (x^{i})$ must satisfy
the same
relations. We keep the same formula for them, namely
\be \coadd(x^{i})=x^{i} \otimes 1 + 1\otimes x ^{i}, \ee
but change to a non-commutative algebra structure in ${\cal A} \otimes
{\cal A}$:
\bea
(1 \otimes x^{i})(x^{j}\otimes 1)&=& B^{ij}_{kl}(x^{k} \otimes 1)(1 \otimes
x^{l}) \nonumber \\ &=&B^{ij}_{kl}x^{k} \otimes  x^{l}     \label{eq:braid}
\eea
where $B$ satisfies the braid relation and commutes with $R$.

     Now let $v$ be a vector in our non-commutative space $V$ (formally,
an element of the dual vector space to the space spanned by the $x^{i}$).
Then \cite{Mdiff} we can define $D_{v}$, the directional derivative along
$v$,
by following the classical idea that the directional derivative of the
function $f$ at the point with coordinates $x$ is obtained from $f(x+y)$
by
picking out the first-order terms in $y$ and evaluating them with $y$
equal to the coordinates of $v$. Writing $d_{v}f=D_{v}f(0)=f_{1}(v)$
where $f_{1}$ is the first-order part of
the function
$f$, this prescription for forming $D_{v}f$ as a function
of $x$
is
\be
D_{v}=(\mbox{id}\otimes d_{v})\coadd (f)
\ee
which is also a good definition of $D_{v}:{\cal A} \rightarrow {\cal A}$
when ${\cal A}$
is
a non-commutative algebra of functions. Exactly as in commutative calculus,
we define $df$ to be the map which associates to each vector $v \in
V$ the directional derivative $D_{v}f$; this is a linear
map $df:
V
\rightarrow {\cal A}$ and can therefore be regarded as an element of ${\cal
A}  \otimes
V^{*}$ which is a subset of ${\cal A} \otimes {\cal A}$ since $V^{*}$
generates ${\cal A}$.
Identifying the coordinates $x^{i}$ with $x^{i} \otimes 1$, we have both
coordinates and differentials in the algebra ${\cal A} \otimes {\cal A}$,
with the relations
(\ref{eq:braid}); the differentials $dx^{i}$ are identified with $1\otimes
x^{i}$, so the relations become
\be
dx^{i}x^{j}=B^{ij}_{kl}x^{k}dx^{l}.   \label{eq:dxx}
\ee
These relations are usually taken as a starting point
for a non-commutative differential
calculus.
\vspace{5mm}
\begin{samepage}
\bc
INVARIANT VECTOR FIELDS
\ec

       Now let us try to follow the classical construction of the Lie
\end{samepage}
algebra of a Lie group as the set of left-invariant vector fields on the
group manifold. Let ${\cal H}$ be a Hopf algebra, which we think of as
analogous
to the function space $C^{\infty}(G)$ of a Lie group $G$. There is a natural
definition of a left-invariant operator $X:{\cal H}\rightarrow {\cal H}$
in terms of
the
coproduct $\Delta$ of ${\cal H}$: $X$ is left-invariant if
\be
\Delta \circ X = (\mbox{id}  \otimes X) \circ \Delta.
\ee
The set $L(H)$ of all such left-invariant operators is closed under operator
multiplication, and is isomorphic as an algebra to the Hopf dual ${\cal
H}^{*}$
by the correspondence
\bea
L(H) \ni X & \mapsto & X_{e}=\epsilon \circ X \in H^{*}\\
H^{*} \ni X_{e} & \mapsto & X = (\mbox{id}  \otimes X_{e}) \circ \Delta
\in L(H)
\eea
where $\epsilon$ is the counit of ${\cal H}$. (Classically, if $X$ is
a differential operator, $X_{e}$ maps $f \in {\cal H}$
to the number {\em Xf}$\, (e)$ where $e$ is the identity of $G$; in
particular,
if $X$ is a vector field then $X_{e}$ is the corresponding tangent vector
at the identity.) If $X_{e} \in H^{*}$ has a coproduct $\mu ^{*}(X_{e})
= \sum X_{(1)e} \otimes X_{(2)e}$, where $\mu$ is the multiplication in
${\cal H}$, then the corresponding $X \in L(H)$ satisfies
\be
X(fg) = \sum (X_{(1)}f)(X_{(2)}g),
\ee
i.e. it is a generalised derivation of ${\cal H}$. Classically,
being a derivation is the defining property of a vector field (i.e. a
first-order differential operator); but the generalised property of derivation
provided by Hopf algebra theory is too general for this definition, for
it includes higher-order differential operators. To capture the notion
of a {\em first-order} differential operator we need more structure than
that of a Hopf algebra on the algebra of functions; this is just what
is provided by Woronowicz's idea of a differential calculus.

      Rearranging Woronowicz's theory a little, we can take a differential
calculus on a Hopf algebra ${\cal H}$ to be a set ${\cal D}$ (of differential
forms)
which is an ${\cal H}$-bimodule and a differential bialgebra, i.e. ${\cal
D}$ has the
following structures:
\begin {enumerate}
\item An \N-grading ${\cal D}=D_{0} \oplus D_{1} \oplus \cdots \mbox{
}$ with $D_{0}={\cal H}$.
We can describe this grading by a map $\nu :{\cal D} \rightarrow {\cal
D}$ such that
$D_{n}=\mbox{
ker}(\nu - n)$.
\item A $\Z_{2}$-grading $\sigma = (-1)^{\nu}$.
\item A differential $d:{\cal D} \rightarrow {\cal H}$ of $\nu$-degree
1, i.e. $d(D_{n})
\subseteq D_{n+1}$ and
\be d(\theta \phi)=(d\theta)\phi + \sigma(\theta)d\phi .
\ee
\item An algebra homomorphism $\widetilde{\Delta}:{\cal D} \rightarrow
{\cal D}$ with $\widetilde{\Delta}|H=\Delta$
and
\be \widetilde{\Delta} \circ d = (d  \otimes 1 + \sigma \otimes d)
\circ \widetilde{\Delta};
\ee
\item $D_{1}$ is generated by $d(H)$ as a left ${\cal H}$-module.
\end {enumerate}
Thus differential forms can be multiplied by functions both on the left
and on the right, and there are relations giving a product of the form
$df.g$ as a sum of products of the form $g' df'$.

        We define a {\em vector field} on ${\cal H}$(relative to the
differential calculus ${\cal D}$) to be an operator
$X:{\cal H} \rightarrow {\cal H}$ which is of the form
$X= \iota _{X} \circ d$ where $\iota _{X}:D_{1} \rightarrow {\cal H}$
is left
${\cal H}$-linear. Such an operator satisfies a deformed version of the
usual
derivation property which defines a vector field classically:
\be X(fg)=f(Xg) + \sum g'(Xf') \mbox{     where    }
df.g=\sum g'df'.
\label{eq: deriv} \ee
It follows from this that the action of a vector field on elements of
${\cal H}$ is determined by its action on the generators of ${\cal H}$.
Left-invariant
vector fields correspond to elements of ${\cal H}^{*}$, so if ${\cal H}$
has a finite
number of generators then the set of left-invariant vector fields is a
finite-dimensional vector space.
\vspace{5 mm}
\bc THE LIE ALGEBRA OF A QUANTUM GROUP \ec

    To define a Lie bracket between left-invariant vector fields, as just
defined, we return to the general Hopf-algebra bracket (\ref{eq: adbrac}).
Between elements $X_{e}$, $Y_{e}$ of the dual Hopf algebra ${\cal H}^{*}$,
this
bracket can be written
\be
\langle [X_{e}, Y_{e}], f\rangle =
\langle X_{e} \otimes Y_{e}, \mbox{ad}(f)\rangle
\label {eq: dualbrac} \ee
where the angle brackets denote the pairing between a vector space and
its dual, and ad:${\cal H} \rightarrow {\cal H} \otimes {\cal H}$ is the
adjoint coaction:
\be \mbox{ad}(f)=\sum f_{(1)}\kappa (f_{(3)}) \otimes f_{(2)}.  \ee
In the classical case, where $f$ is a function of one variable in $G$,
ad$(f)$ is the function of two variables given by ad$(f)(x,y)=f(xyx^{-1})$,
and tangent vectors at the identity like $X_{e}$ are given as elements
of ${\cal H}^{*}$ by
\be X_{e}f=\frac{d}{dt} f(e^{tX})\Bigr|_{t=0}, \nonumber \ee
the definition (\ref{eq: dualbrac}) corresponds to
\be [X_{e}, Y_{e}]f=\frac{d}{ds}\frac{d}{dt}
f(e^{sX}e^{tY}e^{-sX})\Bigr|_{s=t=0}
\nonumber \ee
By the isomorphism between ${\cal H}^{*}$ and $L({\cal H})$, this definition
gives a
Lie bracket between left-invariant differential operators. The classical
fact that the set of tangent vectors is closed under this bracket carries
over to the general case of a Hopf algebra with a differential calculus,
and so do the familiar properties of the Lie bracket:

{\bf Theorem} (Woronowicz \cite{W}).
Let ${\cal H}$ be a Hopf algebra with a differential calculus, and let
$L$ be
the set of left-invariant fields on ${\cal H}$. Then:
\begin{enumerate}
\item If $X$, $Y$ are left-invariant vector fields on ${\cal H}$,
so is $[X,Y]$.
\item There is a braiding $\beta :L \otimes L\rightarrow L \otimes L$ such
that \begin{enumerate}
\item[(a)] $[X,Y]=XY-\mu \circ \beta (X \otimes Y)$ \hfill (quommutator)
\item[(b)] For $T \in L \otimes L$, \hspace{1cm} $\beta (T)=T \Longrightarrow
[T]=0$ \hfill (q-antisymmetry)
\item[(c)] ad$_{L}[X,Y]=\mbox{ad}_{L}X.\mbox{ad}_{L}Y-\mu \circ
(\mbox{ad}_L \otimes \mbox{ad}_{L})\circ \beta (X \otimes Y)$ \hfill (q-Jacobi)
\end{enumerate}
\end{enumerate}
where $\ad_{L}X$ takes $Y \in L$ to $[X,Y]\in L$.
\vspace{5 mm}
\bc THE GOOD BEHAVIOUR OF GL$_{q}(n)$ \ec

When ${\cal H}$ is the quantised function algebra of GL$(n)$, this theory
works
as well as one could possibly hope. In this case ${\cal H}$ is generated
by the
elements of an $n \times n$ matrix $A=(a^{i}_{j})$ with relations
\be
R_{12} A_{1} A_{2} = A_{1} A_{2} R_{12},\mbox{\hspace{1cm}i.e.\hspace{0.5cm}}
R^{ij}_{kl} a^{k}_{m} a^{l}_{n} = a^{i}_{k} a^{j}_{l} R^{kl}_{mn},
\label{eq:RAA} \ee
where the $n^{2}\times n^{2}$ matrix $R$ is diagonalisable with
two eigenvalues $\lambda =q$ and $\mu =-q^{-1}$, so that $(R-q)(R+q^{-1})=0$.
Then a differential calculus on ${\cal H}$ is defined by the commutation
relations
\be
dA_{1} A_{2} = -\mu ^{-1}R_{12}A_{1}dA_{2}R_{12}
\ee
(which are categorically determined \cite{ASdiff} by a differential calculus
on the associated quantum space). This differential calculus yields $n^{2}$
independent left-invariant vector fields and a matrix of $n^{2}$ independent
left-invariant (Maurer-Cartan) forms $\bf{\Omega }=A^{-1}d{\bf
A}$, and hence an $n^{2}$-dimensional Lie algebra with a basis $E^{i}_{j}$
and Lie brackets
\be
[E^{i}_{j},E^{k}_{l}]=E^{i}_{j}E^{k}_{l}-R^{ke}_{jf}R^{fi}_{ga}(R^{\sim
1})^{bg}_{ch}(R^{-1})^{hd}_{el}E^{a}_{b}E^{c}_{d}
\ee
where $R^{-1}$ is the inverse of $R$ as an element of End$(\C^{n}
\otimes \C^{n})$ (acting on contravariant tensors $t^{ij}$),
and $R^{\sim 1}$ is the inverse of the same matrix as an element of
End(End$\C^{n}$) (acting on mixed tensors $t^{i}_{j}$); i.e.
\be
(R^{-1})^{ij}_{kl}R^{kl}_{mn}=\delta^{i}_{m}\delta^{j}_{n}\hspace{1cm}\mbox{
and} \hspace{1cm} (R^{\sim 1})^{ij}_{kl}R^{lm}_{jn}=\delta ^{i}_{n}\delta
_{k}^{m}.
\nonumber \ee
This can be regarded as a bracket on the space of $n\times n$ matrices
$X=x_{i}^{j}E^{j}_{i}$, with
\be
[X,Y]_{i}^{j} = x_{i}^{k}y_{k}^{j} - x_{m}^{n} y_{k}^{l}
 R^{kr}_{ns} R^{sm}_{pi} t^{p}_{q} (R^{-1})^{qj}_{rl}
\ee
where $t^{i}_{j}=(R^{\sim 1})^{ki}_{kj}$ is the quantum trace. This looks
less unpleasant in a graphical notation:\\
\begin{picture}(143.00,90.00)(5,75)
\put(47.00,135.00){\makebox(0,0)[cc]{$[X,Y]$}}
\put(58.64,135.00){\makebox(0,0)[cc]{$=$}}
\put(69.38,125.18){\makebox(0,0)[cc]{$X$}}
\put(69.28,145.18){\makebox(0,0)[cc]{$Y$}}
\put(77.00,135.00){\makebox(0,0)[cc]{$-$}}
\put(101.00,124.00){\makebox(0,0)[cc]{$X$}}
\bezier{164}(85.00,135.00)(84.00,165.00)(93.00,158.00)
\put(102.00,143.00){\makebox(0,0)[cc]{$Y$}}
\bezier{64}(95.00,151.00)(103.00,157.00)(103.00,163.00)
\bezier{16}(93.00,158.00)(94.00,157.00)(96.00,155.00)
\put(16.00,90.00){\makebox(0,0)[rc]{where}}
\put(38.00,89.00){\makebox(0,0)[cc]{$=$}}
\put(43.00,89.00){\makebox(0,0)[lc]{$R^{ij}_{kl}$,}}
\put(23.00,93.00){\makebox(0,0)[cc]{$i$}}
\put(23.00,85.00){\makebox(0,0)[cc]{$k$}}
\put(32.00,85.00){\makebox(0,0)[cc]{$l$}}
\put(32.00,93.00){\makebox(0,0)[cc]{$j$}}
\put(64.00,85.00){\makebox(0,0)[cc]{$k$}}
\put(73.00,85.00){\makebox(0,0)[cc]{$l$}}
\put(64.00,94.00){\makebox(0,0)[cc]{$i$}}
\put(73.00,94.00){\makebox(0,0)[cc]{$j$}}
\put(79.00,89.00){\makebox(0,0)[cc]{$=$}}
\put(84.00,89.00){\makebox(0,0)[lc]{$(R^{-1})^{ij}_{kl}$}}
\put(108.00,89.00){\makebox(0,0)[cc]{and}}
\bezier{76}(127.00,93.00)(130.00,103.00)(132.00,95.00)
\bezier{52}(127.00,85.00)(128.00,78.00)(132.00,83.00)
\put(134.00,84.00){\makebox(0,0)[cc]{$i$}}
\put(134.00,93.00){\makebox(0,0)[cc]{$j$}}
\put(138.00,89.00){\makebox(0,0)[cc]{$=$}}
\put(143.00,89.00){\makebox(0,0)[lc]{$t^{i}_{j}$.}}
\bezier{80}(95.00,151.00)(89.00,141.00)(95.00,135.00)
\bezier{16}(99.00,131.00)(101.00,129.00)(101.00,128.00)
\bezier{40}(96.00,132.00)(102.00,137.00)(102.00,139.00)
\bezier{20}(101.00,147.00)(100.00,150.00)(99.00,151.00)
\put(24.00,86.00){\line(1,1){7.00}}
\put(65.00,93.00){\line(1,-1){7.00}}
\bezier{32}(127.00,85.00)(126.00,90.00)(127.00,93.00)
\bezier{72}(96.00,132.00)(90.00,124.00)(96.00,118.00)
\bezier{60}(96.00,118.00)(103.00,112.00)(103.00,106.00)
\bezier{180}(85.00,135.00)(87.00,104.00)(97.00,114.00)
\bezier{20}(100.00,118.00)(102.00,121.00)(101.00,121.00)
\put(47.00,106.00){\line(0,1){25.00}}
\put(47.00,138.00){\line(0,1){25.00}}
\put(69.00,106.00){\line(0,1){16.00}}
\put(69.00,129.00){\line(0,1){13.00}}
\put(69.00,149.00){\line(0,1){14.00}}
\bezier{12}(24.00,93.00)(25.00,92.00)(26.00,91.00)
\bezier{12}(31.00,86.00)(30.00,87.00)(29.00,88.00)
\bezier{12}(72.00,93.00)(71.00,92.00)(70.00,91.00)
\bezier{12}(65.00,86.00)(66.00,87.00)(67.00,88.00)
\end{picture}

\bc . . . BUT O$_{q}(n)$ WON'T PLAY BALL   \ec

 Now let ${\cal H}$ be the quantised function algebra of GO$(n)$, the
orthogonal
group together with dilatations in $n$ dimensions. Just like the case
of GL$(n)$, this ${\cal H}$ has $n^{2}$ generators $a^{i}_{j}$ forming
a matrix
$A$, subject to relations $R_{12}A_{1}A_{2}=A_{1}A_{2}R_{12}$, but now
the $n^{2}\times n^{2}$ matrix
$R$ has three eigenvalues:
\bea
\lambda _{+}=q;&\quad\mbox{eigenspace } E_{+}&\mbox{with  dim} E_{+}
={\scriptstyle \frac{1}{2}} n(n+1)-1 \nonumber \\
\lambda_{-}=-q^{-1};&\quad\mbox{eigenspace } E_{-}&\mbox{with  dim}E_{-}
= {\scriptstyle \frac{1}{2}} n(n-1) \nonumber \\
\lambda _{0}=q^{1-n}; & \quad \mbox{eigenspace } E_{0} &
\mbox{with  dim}E_{0}=1
\nonumber
\eea
The one-dimensional eigenspace $E_{0}\subset \C^{n} \otimes \C
^{n}$ is spanned by the quantum metric $g^{ij}$ (not a symmetric
tensor). We have projectors $\Pi _{+}$, $\Pi _{-}$, $\Pi _{0}$ onto the
three eigenspaces, with
\be
(\Pi _{0})^{ij}_{kl}=\alpha g^{ij}g_{kl} \nonumber
\ee
where $\alpha $ is a scalar and $g_{ij}$ is the inverse of $g^{ij}$.
In terms of these projectors, the relations can be written
\be
(\Pi _{0}+\Pi _{+})A_{1}A_{2}\Pi _{-}=\Pi _{-}A_{1}{\bf
A}_{2}(\Pi _{0}+\Pi _{+})=0,
\ee
which is a $q$-deformation of the statement that the matrix elements of
$A$ commute, and
\be
\Pi _{0}A_{1}A_{2}\Pi _{+}=\Pi _{+}A_{1}A_{2}\Pi
_{0}=0,
\ee
which is a deformation of the orthogonality equation
 $g^{kl}a^{i}_{k}a^{j}_{l}=Qg^{ij}$.
In the quantum case the scalar $Q$  is given by
\be
\Pi _{0}A_{1}A_{2}\Pi _{0}=Q\Pi _{0} \label{eq:Q}
\ee
and commutes with all the matrix elements $a^{i}_{j}$.

 Suppose that in an attempt to construct a differential calculus on the
algebra ${\cal H}$, we assume the same commutation relations as in the
case of
GL$(n)$:
\be
dA_{1}A_{2}=qR_{12}A_{1}dA_{2}R_{12}
\label{eq:AdA} \ee
Then applying the external derivative $d$ to the relations
$RA_{1}A_{2}=A_{1}A_{2}R$ gives
\be
[R, 1 \otimes \Omega + R(1 \otimes \Omega)R]=0
\ee
where $\Omega=A^{-1}dA$ is the matrix of left-invariant
1-forms, which we expect to be antisymmetric. But from the above equation
we get both
\be
\Pi _{0}(1 \otimes \Omega )\Pi _{+}=0,
\ee
 which says that the traceless part of $\Omega$ is q-antisymmetric, and
\be
\Pi _{0}(1 \otimes \Omega)\Pi _{-}=0,
\ee
which says that $\Omega$ is q-symmetric. These conditions leave
only one independent left-invariant 1-form in the matrix $\Omega$.

 On the other hand, Carow-Watamura et al \cite{CSSW} found a bicovariant
differential calculus on ${\cal H}$ in which all $n^{2}$ 1-forms in the
matrix
$\Omega$ were independent.

 Schm\"{u}dgen and Sch\"{u}ler \cite{SS1,SS2} have considered the general
theory
of a bicovariant differential calculus on the algebra ${\cal H}$. We have
$n^{2}$
left-invariant 1-forms $\omega ^{i}_{j}=\kappa (a^{i}_{k})da^{k}_{j}$;
we can use the quantum metric to lower an index and obtain
$\omega _{ij}=g_{ik}\omega ^{k}_{j}$. It follows from the fact that the
$a^{i}_{j}$ generate ${\cal H}$
that the $\omega _{ij}$ span the space of left-invariant 1-forms. The
general theory of Woronowicz \cite{W} then gives commutation relations
between the $\omega ^{i}_{j}$ and the matrix elements $a^{i}_{j}$ of the
form
\be
\omega _{ij}a^{k}_{l}=a^{k}_{p}T^{pmn}_{ijl}\omega _{mn}
\label{eq:AOmega } \ee
where $T$ is a numerical tensor. The requirement of bicovariance is that
there should be left and right coactions
\be
\delta _{\mbox{{\footnotesize L}}}(da^{i}_{j})=a^{i}_{k} \otimes da^{k}_{j},
\mbox{\hspace{1cm}}
\delta_{\mbox{{\footnotesize R}}}(da^{i}_{j})=da^{i}_{k} \otimes a^{k}_{j}
\nonumber \ee
\be \mbox{i.e. \hspace{1cm}} \delta _{\mbox{{\footnotesize L}}}
(\omega_{ij})=1\otimes \omega _{ij},
\hspace{1cm} \delta_{\mbox{{\footnotesize R}}} (\omega_{ij}) = \omega_{kl}
\otimes a^{k}_{i}a^{l}_{j} \nonumber
\ee
which are ${\cal H}$-bimodule homomorphisms, i.e. are consistent with
the coproducts
of matrix elements and the commutation relations between differentials
and matrix elements. This leads to
\be
T_{123}A_{1}A_{2}A_{3}=A_{1}A_{2}A_{3}T_{123}
\ee
if we assume that the 1-forms $\omega_{ij}$ are all independent over
${\cal H}$. There are two further consistency conditions. By considering
the
product $\Omega_{1}A_{2}A_{3}$ and requiring consistency between
the commutation relations (\ref{eq:RAA}) and (\ref{eq:AOmega }) (where
``consistency" means independence of the 1-forms $\omega _{ij}$ over ${\cal
H}$), we are led to
\be
R_{12}T_{234}T_{123}=T_{234}T_{123}R_{34}.
\ee
Schm\"{u}dgen and Sch\"{u}ler show that this requires $T$ to be of the
form
\be
T_{123}=X_{23}^{-1}R_{12}^{-1}R_{23}X_{12}.      \label{eq: defX}
\ee
Finally, differentiating $RA_{1}A_{2}=A_{1}A_{2}R$ and again requiring
that the $\omega _{ij}$ should be independent, they show that $X$ must
be
\be           X=R-\lambda _{0}^{-1}              \ee
which gives precisely the differential calculus of {\em al}.\ {\em et} Weich
\cite{CSSW}.

          But the assumptions in this argument are unreasonable; the $\omega
_{ij}$ ought not to be independent. This is particularly evident in the
last step: the relations $RA_{1}A_{2}=A_{1}A_{2}R$ include a version
of the orthogonality condition on $A$, and differentiating this ought
to lead to a version of the antisymmetry of $\omega _{ij}$. However, making
the assumptions more reasonable (in classical terms) leaves the conclusion
the same:

{\bf Theorem}. Let ${t_{\alpha }^{ij}}$ be a set of independent
$q$-antisymmetric tensors, i.e. a basis for the image of $\Pi _{-}$, and
suppose that the 1-forms $t_{\alpha }^{ij}\omega _{ij}$ are independent
over ${\cal H}$. Then the matrix $X$ in (\ref{eq: defX}) is $X=R-\lambda
_{0}^{-1}$
and all the $\omega _{ij}$ are independent.
\vspace{5 mm}
\bc THE QUANTUM SPHERE \ec

Another place where we might expect to find a set of vector fields whose
Lie brackets give an orthogonal Lie algebra is on the quantum sphere,
which is a homogeneous space for the quantum orthogonal group. On the
sphere, unlike the group, a differential calculus can be defined which
is a genuine deformation of the classical one.

Homogeneous spaces can be constructed for any Hopf algebra ${\cal H}$
with matrix
comultiplication generated by matrix elements $a^{i}_{j}$ satisfying relations
(\ref{eq:RAA}) with a diagonalisable $R$-matrix
\be
R=\lambda _{1}\Pi _{1}+ \cdots + \lambda _{r}\Pi _{r}
\ee
where the $\Pi _{i}$ are projectors onto the eigenspaces  $E_{i}$ of $R$.
For
each
of these eigenspaces, or for any subset $K=\{ k_{1},\ldots ,k_{p}\}$ of
them, we can form a quantum space $S_{K}$ with generators $x^{i}$ and
relations
\be
\Pi_{k}\x_{1}\x_{2}=0\mbox{\hspace{1cm}for each } k \in
K
\label{eq:qspace} \ee
The space $S_{K}$ then admits a coaction of the bialgebra ${\cal H}$,
\be
\delta _{K}:S_{K}\rightarrow H \otimes S_{K} \mbox{\hspace{1cm} with
\hspace{1cm}} x^{i} \mapsto a^{i}_{j} \otimes x^{j},
\label{eq:coaction} \ee
that is to say the relations (\ref{eq:qspace}) are preserved by this map.
We can also consider spaces defined by an equation like (\ref{eq:qspace}) but
with a non-zero right-hand side in the form of a numerical
tensor {\bf t}$_{k} \in E_{k} \subset \C^{n} \otimes \C^{n}$. To obtain
a coaction on such a space it will be necessary to add further relations to
(\ref{eq:RAA}). In the case of the quantum orthogonal group we have
\be
R=q\Pi _{+}-q^{-1}\Pi _{-}+\lambda _{0}\Pi _{0}
\ee
and the quantum sphere is given by the relations
\be
\hspace{1cm}\Pi _{-}\x_{1}\x_{2}=0 \mbox{\hspace{1cm} and \hspace{1cm}} \Pi
_{0}\x_{1}\x_{2}=\rho \mbox{{\bf g}}
\ee
where $\rho $ is the radius of the sphere. The first equation, a deformation
of the statement that the coordinates commute, defines a $q$-orthogonal
flat space; the second is the equation of a sphere in this space. They
can be combined in the single equation
\be
R^{ij}_{kl}x^{k}x^{l}=qx^{i}x^{j}+\rho (\lambda _{0}-q)g^{ij}
\label{eq:sphere} \ee
If $\rho \neq 0$ the relations (\ref{eq:RAA}) must be supplemented by
\be g^{kl}a_{k}^{i}a_{l}^{j}=g^{ij} \ee
i.e. $Q=1$ in (\ref{eq:Q}).
We will use ${\cal F}(S)$ to denote the algebra of functions on the sphere,
i.e. the algebra generated by the $x^{i}$ with the relations
(\ref{eq:sphere}). We define a differential calculus ${\cal D}(S)$
on the sphere by means of the commutation relations
\be
\mbox{\bf dx}_{1}\x_{2}=qR_{12}\x_{1}\mbox{\bf dx}_{2}
\label{eq:xdx} \ee
and the relations between the $dx^{i}$ obtained by differentiating this.
Differentiating the sphere equation (\ref{eq:sphere}) then yields
\be
\Pi _{0}\x_{1}\mbox{\bf dx}_{2}=0,\hspace{1cm} \mbox{i.e.} \hspace{1cm}
g_{ij}x^{i}dx^{j}=0
\ee
as we would expect classically. Thus the number of independent monomials
in coordinates and differentials is the same as for the classical sphere.

 We define a vector field on the sphere to be a left ${\cal F}(S)$-linear
map $X:{\cal F}(S)\rightarrow {\cal F}(S)$ which is obtained from a map
$\iota _{X}:{\cal D}(S)\rightarrow {\cal F}(S)$ by $X=\iota _{X}\circ d$.
Then $X$ has a derivation property,
and is determined by its action on the generators $x^{i}$ by
\be
X(\x_{1} \ldots \x_{n})=(1+qR_{n-1,n}+\cdots +q^{n-1}R_{12}R_{23}\cdots
R_{n-1,n})\x_{1}\cdots \x_{n-1}X(\x_{n-1}).
\label{eq:Sderiv} \ee
\newpage
 \bc
INFINITESIMAL  ROTATIONS
\ec

Even in the absence of a notion of a tangent vector at the identity of
the orthogonal group, we can distinguish those vector fields on the quantum
sphere which are transmitted from differential operators at the identity
of the group by means of the coaction $\delta $ of (\ref{eq:coaction}).
Such an {\em infinitesimal rotation} is a vector field whose action on
${\cal F}(S)$ is given by
\be
\mbox{{\em Xf}} = (\xi  \otimes \mbox{id}) \delta (f) \hspace{1cm} \mbox{for
some }\xi  \in H^{*}.
\ee
The linearity of the coaction then means that the action of $X$ on the
coordinates must be linear:
\be
X(x^{i})=m^{i}_{j}x^{j}
\label{eq:XM}\ee
for some numerical matrix $\mbox{{\bf M}}=(a^{i}_{j})$. The derivation
equation
(\ref{eq:Sderiv}) together with the sphere equation (\ref{eq:sphere})
then
imply that {\bf M} must satisfy
\be
\Pi _{0}(1 \otimes \mbox{\bf M})\Pi _{+}=0
\ee
which is a deformation of the statement that {\bf M} (with its trace removed)
is antisymmetric with respect to the quantum metric $g$. This is what
one would expect classically. However, in the quantum case there is an
extra condition
\be
\rho \Pi _{0}(1 \otimes \mbox{\bf M})\Pi _{-}=0
\ee
which, if $\rho \not= 0$, implies that {\bf M} is q-symmetric as well
as q-antisymmetric, and therefore {\bf M} must be a multiple of the identity
matrix. Thus infinitesimal rotations exist only on the null-sphere.
\vspace{5 mm} \bc
THE LIE BRACKET OF VECTOR FIELDS ON A QUANTUM HOMOGENEOUS SPACE
\ec
 We have just seen that a homogeneous space of a quantum group may have
a satisfactory notion of vector field which the group itself lacks. On
the other hand, on a quantum space with a differential calculus there
is no natural definition of the Lie bracket of vector fields; the coaction
of the group can remedy this deficiency.

 Let ${\cal F}$ be a (non-commutative) function algebra with a differential
calculus ${\cal D}$, and let ${\cal H}={\cal F}(G)$ be a Hopf algebra
coacting
on ${\cal F}$, so that there is an algebra homomorphism $\delta :{\cal
F}\rightarrow {\cal H} \otimes {\cal F}$. As on the quantum sphere, we
define
an {\em infinitesimal generator} of the coaction to be a map $X:{\cal
F}\rightarrow {\cal F}$ which is both a vector field, so that $X=\iota
_{X}\circ d$ for some map $\iota _{X}:{\cal D}\rightarrow {\cal F}$, and
is obtained from the coaction by
\be
X=(\xi  \otimes \mbox{id})\circ \delta \hspace{1cm} \mbox{for some }\xi
\in H^{*}.
\ee
If $X$, $Y$ are two such infinitesimal generators obtained from $\xi ,\eta
\in H^{*}$, we can form the adjoint Lie bracket $[\xi ,\eta ]$ by (\ref{eq:
dualbrac})
and hence define the Lie bracket of $X$ and $Y$ by
\be
\;[X,Y]=([\xi ,\eta ] \otimes \mbox{id})\circ \delta
\ee
It can be shown that $[X,Y]$ is also a vector field on ${\cal F}$. However,
in general, a given infinitesimal generator $X$ is not obtained from a
unique $\xi \in H^{*}$, and $[X,Y]$ will depend on the choice of $\xi$
and $\eta $.
\vspace{5 mm}
\bc
THE  SLIPPERY  PATH  FROM  SPHERE  TO  GROUP
\ec
 On the Sunday in the middle of the Karpacz school, some of us climbed
Snie\.{z}ka and discovered how slippery uphill paths can be. The attempt
to lift vector fields from the quantum sphere to the orthogonal group
is reminiscent of that walk.

 Suppose $X$ is an infinitesimal rotation of the quantum sphere determined
by $\xi \in H^{*}$. Then the action of $X$ on coordinates is given by
a matrix {\bf M} according to (\ref{eq:XM}), and this implies that the
value
of $\xi $ on the generators $a^{i}_{j}$ of ${\cal H}={\cal F}(G)$ is also
given
by {\bf M} according to
\be
\langle \xi ,a^{i}_{j}\rangle =m^{i}_{k}a^{k}_{j}
\nonumber \ee
To define $\xi $ completely, we need to specify $\langle \xi ,f\rangle
$ for all polynomials $f \in {\cal F}(G)$.

 If $\xi =\widetilde{X}_{e}$ where $\widetilde{X}$ is a vector
field on
$G$ with respect to some differential calculus on ${\cal F}(G)$, then
\bea
X(\x_{1}\x_{2})&=&\iota _{X}(\x_{1}\mbox{\bf dx}_{2}+
\mbox{\bf dx}_{1}\x_{2})
\nonumber \\
=\langle \xi ,\pi A_{1}\pi A_{2}\rangle \x_{1}\x_{2}&=&
\iota _{\stackrel{\sim}{X}}(A_{1}dA_{2}+dA_{1}A_{2})\x_{1}\x_{2}
\nonumber \eea
But this forces the differential calculus on ${\cal F}(G)$ to be the
impoverished one that we first considered, with relations (\ref{eq:AdA}):

{\bf Theorem}. The only differential calculi on ${\cal F}(S)$ and ${\cal
F}(G)$
which are compatible in the above sense are defined by the relations
\bea
\x_{1}\x_{2}&=&qR_{12}\x_{1}d\x_{2} \nonumber \\
dA_{1}A_{2}&=&qR_{12}A_{1}dA_{2}R_{12}\nonumber
\eea
In terms of the general analysis of Schm\"{u}dgen and Sch\"{u}ler, this
differential calculus on the group has
$T_{123}=X^{-1}_{23}R^{-1}_{12}R_{23}X_{12}$ with $X=1$.
\vspace{5 mm} \bc PRESS ON REGARDLESS \ec
 Suppose we just ignore the unsatisfactory features of the differential
calculus defined by (\ref{eq:AdA}), and develop its formal consequences.
It gives a derivation property for vector fields $\widetilde{X}$
on the
quantum orthogonal group which makes it possible to evaluate $\widetilde
{X}_{e}$
on products of matrix elements; for example,
\be
\langle \stackrel{\sim}{X}_{e}, A_{1}A_{2}A_{3}\rangle =
\langle \stackrel{\sim}{X}_{e},
A_{3}\rangle +R_{23}\langle \stackrel{\sim}{X}_{e}, A_{3}\rangle R_{23}
+ R_{12}R_{23}\langle \stackrel{\sim}{X}_{e}, A_{3}\rangle R_{23}R_{12}
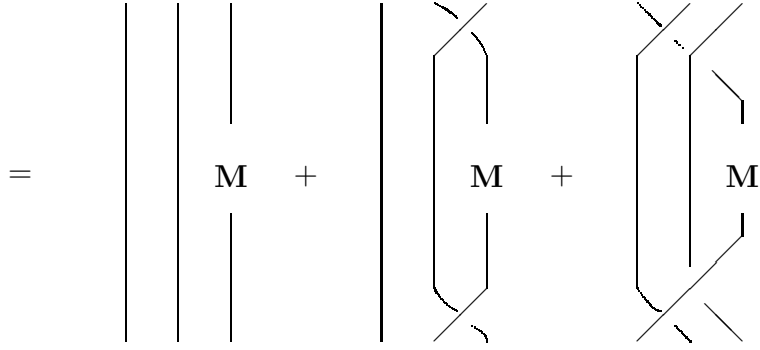
\ee
\begin{picture}(123.00,50.00)(0,85)
\put(27.00,112.00){\makebox(0,0)[cc]{$=$}}
\put(41.00,90.00){\line(0,1){45.00}}
\put(48.00,90.00){\line(0,1){45.00}}
\put(65.00,112.00){\makebox(0,0)[cc]{$+$}}
\put(55.00,90.00){\line(0,1){17.00}}
\put(55.00,135.00){\line(0,-1){16.00}}
\put(55.00,112.00){\makebox(0,0)[cc]{{\bf M}}}
\put(75.00,90.00){\line(0,1){45.00}}
\put(89.00,135.00){\line(-1,-1){7.00}}
\put(82.00,128.00){\line(0,-1){31.00}}
\put(82.00,90.00){\line(1,1){7.00}}
\put(89.00,97.00){\line(0,1){10.00}}
\put(89.00,119.00){\line(0,1){9.00}}
\put(89.00,112.00){\makebox(0,0)[cc]{{\bf M}}}
\put(99.00,112.00){\makebox(0,0)[cc]{$+$}}
\put(116.00,135.00){\line(-1,-1){7.00}}
\put(109.00,128.00){\line(0,-1){31.00}}
\put(109.00,90.00){\line(1,1){14.00}}
\put(123.00,104.00){\line(0,1){3.00}}
\put(123.00,135.00){\line(-1,-1){7.00}}
\put(116.00,128.00){\line(0,-1){28.00}}
\put(123.00,90.00){\line(-1,1){5.00}}
\put(114.00,130.00){\line(1,-1){1.00}}
\put(119.00,126.00){\line(1,-1){4.00}}
\put(123.00,122.00){\line(0,-1){3.00}}
\put(123.00,112.00){\makebox(0,0)[cc]{{\bf M}}}
\bezier{16}(89.00,128.00)(88.00,130.00)(87.00,131.00)
\bezier{16}(85.00,133.00)(84.00,134.00)(82.00,135.00)
\bezier{16}(82.00,97.00)(83.00,95.00)(85.00,94.00)
\bezier{12}(87.00,92.00)(89.00,91.00)(89.00,90.00)
\bezier{16}(109.00,135.00)(111.00,133.00)(112.00,132.00)
\bezier{4}(114.00,130.00)(114.00,130.00)(115.00,129.00)
\bezier{20}(109.00,97.00)(111.00,94.00)(112.00,94.00)
\bezier{12}(114.00,92.00)(115.00,91.00)(116.00,90.00)
\end{picture}

This makes sense (when multiplied by $\x_{1}\x_{2}\x_{3}$) even if the
differential calculus doesn't, and gives a Lie algebra between
antisymmetric $n\times n$ matrices $\mbox{{\bf X}}=(x^{i}_{j})$ which
is best presented graphically and in terms of $x^{ij}=g^{ik}x_{k}^{j}$:\\
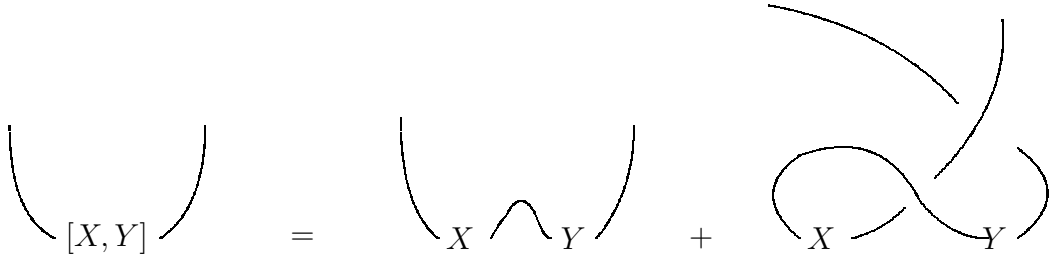
\vspace{5mm}
\begin{picture}(149.00,50.00)(0,110)
\put(20.00,118.00){\makebox(0,0)[cc]{$[X,Y]$}}
\bezier{72}(27.00,118.00)(33.00,122.00)(33.00,133.00)
\bezier{76}(13.00,118.00)(7.00,121.00)(7.00,133.00)
\put(46.00,118.00){\makebox(0,0)[cc]{$=$}}
\bezier{72}(59.00,134.00)(59.00,122.00)(64.00,118.00)
\put(67.00,118.00){\makebox(0,0)[cc]{$X$}}
\bezier{36}(73.00,121.00)(75.00,125.00)(77.00,121.00)
\bezier{16}(77.00,121.00)(78.00,118.00)(79.00,118.00)
\put(82.00,118.00){\makebox(0,0)[cc]{$Y$}}
\bezier{68}(85.00,118.00)(90.00,124.00)(90.00,133.00)
\put(99.00,118.00){\makebox(0,0)[cc]{$+$}}
\bezier{72}(112.00,118.00)(105.00,124.00)(112.00,129.00)
\bezier{92}(112.00,129.00)(123.00,133.00)(128.00,123.00)
\bezier{44}(128.00,123.00)(132.00,118.00)(137.00,118.00)
\put(138.00,118.00){\makebox(0,0)[cc]{$Y$}}
\bezier{80}(141.00,118.00)(149.00,124.00)(141.00,130.00)
\put(115.00,118.00){\makebox(0,0)[cc]{$X$}}
\bezier{120}(133.00,136.00)(124.00,146.00)(108.00,149.00)
\bezier{32}(119.00,118.00)(123.00,119.00)(126.00,122.00)
\bezier{100}(130.00,126.00)(140.00,136.00)(139.00,147.00)
\bezier{16}(71.00,118.00)(72.00,120.00)(73.00,121.00)
\end{picture}\\
(classically, $[X,Y]=XY+Y^{\top}X$). This has surprisingly good properties:

\vspace{5mm}
{\bf Theorem}. Let $T$ be the set of covariant tensors $X=(x^{ij})$,
and let $L$ be the subset satisfying
$\Pi _{+}(X)=0$. Define the Lie bracket $[X,Y]$ between two such tensors
by the above diagram. Then we have
\begin{enumerate}
\item {\bf Closure}
\be    X,Y \in L \Longrightarrow [X,Y] \in L  \ee
\item {\bf Braiding} There is an operator $\beta :M \otimes M\rightarrow
M \otimes M$ satisfying the braid relation $\beta _{12}\beta _{23}\beta
_{12}=\beta _{23}\beta _{12}\beta _{23}$, such that
\be [X,Y]=X \bullet Y-\bullet\beta (X \otimes Y) \ee
where $(X \bullet Y)^{ij}=x^{ik}g_{kl}y^{lj}$, defining $\bullet : M \otimes
M \rightarrow M$, and $\bullet \beta : M \otimes M \rightarrow M$ is the
composition of $\beta$ and $\bullet$.
\item {\bf Antisymmetry} If $T=\sum X_{i} \otimes Y_{i}$ satisfies $\beta
(T)=0$, then $\sum[X_{i},Y_{i}]=0$.
\end{enumerate}
However, the braiding $\beta$ is not defined purely in terms of the
putative Lie algebra $L$ itself; and there is no Jacobi identity
for this bracket. The braiding
$\beta $ is shown in the following diagram:\\
\begin{picture}(130.00,75.00)(0,70)
\put(37.00,117.00){\makebox(0,0)[cc]{$=$}}
\put(48.00,117.00){\makebox(0,0)[cc]{$-\lambda_{0}^{-1}$}}
\bezier{72}(48.00,86.00)(50.00,95.00)(58.00,92.00)
\bezier{72}(58.00,92.00)(67.00,88.00)(69.00,96.00)
\bezier{32}(59.00,90.00)(58.00,85.00)(55.00,84.00)
\bezier{84}(60.00,94.00)(63.00,105.00)(71.00,99.00)
\bezier{60}(71.00,99.00)(78.00,96.00)(83.00,102.00)
\put(83.00,102.00){\line(1,1){33.00}}
\put(79.00,113.00){\line(1,1){28.00}}
\bezier{84}(97.00,109.00)(102.00,99.00)(111.00,96.00)
\bezier{56}(115.00,94.00)(115.00,86.00)(119.00,81.00)
\bezier{84}(119.00,93.00)(130.00,88.00)(127.00,80.00)
\put(93.00,131.00){\line(-1,1){13.00}}
\put(84.00,123.00){\line(-1,1){12.00}}
\put(23.00,117.00){\makebox(0,0)[cc]{$\beta$}}
\put(0.00,77.00){\makebox(0,0)[lc]{since}}
\bezier{112}(106.00,119.00)(114.00,113.00)(115.00,95.00)
\bezier{56}(79.00,113.00)(73.00,109.00)(71.00,102.00)
\end{picture}

\vspace{1cm}
\begin{picture}(153.00,90.00)(15,50)
\bezier{72}(15.00,85.00)(8.00,91.00)(15.00,96.00)
\bezier{92}(15.00,96.00)(26.00,100.00)(31.00,90.00)
\bezier{44}(31.00,90.00)(35.00,85.00)(40.00,85.00)
\put(41.00,85.00){\makebox(0,0)[cc]{$Y$}}
\bezier{80}(44.00,85.00)(52.00,91.00)(44.00,97.00)
\put(18.00,85.00){\makebox(0,0)[cc]{$X$}}
\bezier{120}(36.00,103.00)(27.00,113.00)(11.00,116.00)
\bezier{32}(22.00,85.00)(26.00,86.00)(29.00,89.00)
\bezier{100}(33.00,93.00)(43.00,103.00)(42.00,114.00)
\put(60.00,101.00){\makebox(0,0)[cc]{$=$}}
\put(71.00,101.00){\makebox(0,0)[cc]{$\lambda_{0}^{-1}$}}
\put(74.00,67.00){\makebox(0,0)[cc]{$X$}}
\bezier{72}(71.00,70.00)(73.00,79.00)(81.00,76.00)
\bezier{72}(81.00,76.00)(90.00,72.00)(92.00,80.00)
\bezier{32}(82.00,74.00)(81.00,69.00)(78.00,68.00)
\bezier{84}(83.00,78.00)(86.00,89.00)(94.00,83.00)
\bezier{60}(94.00,83.00)(101.00,80.00)(106.00,86.00)
\put(106.00,86.00){\line(1,1){33.00}}
\put(102.00,97.00){\line(1,1){28.00}}
\bezier{84}(120.00,93.00)(125.00,83.00)(134.00,80.00)
\bezier{56}(138.00,78.00)(138.00,70.00)(142.00,65.00)
\put(146.00,63.00){\makebox(0,0)[cc]{$Y$}}
\bezier{84}(142.00,77.00)(153.00,72.00)(150.00,64.00)
\put(116.00,115.00){\line(-1,1){13.00}}
\put(107.00,107.00){\line(-1,1){12.00}}
\bezier{228}(103.00,128.00)(86.00,148.00)(117.00,148.00)
\bezier{276}(130.00,125.00)(150.00,148.00)(111.00,148.00)
\bezier{108}(128.00,102.00)(136.00,93.00)(138.00,78.00)
\bezier{56}(102.00,97.00)(97.00,93.00)(95.00,86.00)
\end{picture}

\end{document}